\documentclass[12pt]{iopart}
\usepackage{iopams}  
\usepackage{epsfig}
\eqnobysec
\newcommand{\be}{\begin{equation}}
\newcommand{\ee}{\end{equation}}
\newcommand{\ben}{\begin{equation*}}
\newcommand{\een}{\end{equation*}}
\newcommand{\bea}{\begin{eqnarray}}
\newcommand{\eea}{\end{eqnarray}}
\newcommand{\nn}{\nonumber}

\begin{document}

\title[Casimir Repulsion]{Repulsive Casimir and Casimir-Polder Forces}

\author{Kimball A Milton, E K Abalo,  Prachi Parashar and Nima Pourtolami}

\address{Homer L. Dodge Department of Physics and Astronomy,
University of Oklahoma, Norman, OK 73019 USA}

\author{Iver Brevik and 
Simen \AA\ Ellingsen}
\address{Department of Energy and Process Engineering, Norwegian University
of Science and Technology, N-7491 Trondheim, Norway}

\ead{milton@nhn.ou.edu}
\begin{abstract}
Casimir and Casimir-Polder repulsion have been known for more
than 50 years.  The general ``Lifshitz'' configuration of parallel
semi-infinite dielectric slabs permits repulsion if they are separated
by a dielectric fluid that has a value of permittivity that is
intermediate between those of the dielectric slabs.  This was indirectly
confirmed in the 1970s, and more directly by Capasso's group recently.
It has also been known for many years that  electrically and magnetically
polarizable bodies can experience a repulsive quantum vacuum force.
More amenable to practical application are situations where repulsion
could be achieved between ordinary conducting and dielectric bodies in
vacuum.  The status of the field of Casimir repulsion with emphasis on
some recent developments will be surveyed.  Here, stress will be placed
on analytic developments, especially of Casimir-Polder (CP) interactions between
anisotropically polarizable atoms, and CP interactions between 
anisotropic atoms and bodies that also exhibit
anisotropy, either because of anisotropic constituents, or because of
geometry.  Repulsion occurs for wedge-shaped and cylindrical conductors,
provided the geometry is sufficiently asymmetric, that is, either the
wedge is sufficiently sharp or the atom is sufficiently far from the cylinder.

\end{abstract}

\maketitle

\section{Introduction}
Van der Waals forces are generally regarded as attractive, and the same holds
when we pass to the retarded regime of larger distances, where we have
Casimir and Casimir-Polder forces.  But it has long been
recognized that repulsive effects can be achieved.  For example, if two
dielectric bodies are separated by a medium with permittivity of intermediate value, 
the two bodies are repelled because the (fluid) medium is pulled
into the gap between the bodies.  The theory of this was first worked out
by Dzyaloshinskii,  Lifshitz, and Pitaevskii in 1961 \cite{lifshitz61}.
It was accurately verified experimentally by Sabisky and Anderson in 1973 \cite{sabisky}
who realized exactly this configuration (fluoride substrate, helium film, helium vapor).
Recently, much publicity has been given to the explicit observation of
repulsion in this sense by Munday, Capasso, and Parsegian \cite{capasso09}.
We will discuss this type of repulsion in \sref{sec:lifshitz}.

In 1968 Boyer \cite{boyer} showed that, contrary to expectation \cite{Casimir53},
 the quantum self-energy of a perfectly conducting spherical shell of zero
thickness is positive, that is, repulsive in a naive sense.  The recent progress
in understanding the general systematics of the geometry-dependence of
Casimir self-stresses is reviewed in \sref{sec:boyer}.  But this type of effect
is primarily of fundamental interest, because these self-energies or stresses
seem largely inaccessible to observation.  

More interesting is the effect, also discovered by Boyer \cite{boyer74},
of the repulsion of a perfectly conducting plate by a parallel plate
having perfect magnetic conductivity.  This is a straightforward generalization
of the Lifshitz theory for dielectric slabs.  Although it might be thought that
such an effect could be mimicked by metamaterials, it seems now unlikely to be
achievable in practice.  A brief review of this subject is given in \sref{sec:magnetic}.

Metamaterials have been famously shown to exhibit negative refraction 
(perfect lens) behaviour in a limited frequency range \cite{pendry00}. 
If, as a thought experiment, materials which are perfect lenses at all 
frequencies could be manufactured, a repulsive Casimir force \cite{leonhardt07} 
could result. Superlens response at all frequencies is inconsistent with the 
Kramers-Kronig relations, however, almost certainly precluding this scheme. 
A more reasonable set-up is the Casimir-Polder force on an excited atom 
near a superlens \cite{sambale08,sambale09}, since such an atom has a 
resonant interaction at a single frequency. Excited atoms are moreover 
subject to forces whose sign oscillates in space away from any (regular) 
surface, as has been known for a long time \cite{barton70}. However, atoms 
stay excited only for very short times, making this effect difficult to 
utilize. In this respect ultracold molecules \cite{ellingsen09} or 
Rydberg atoms \cite{crosse10} would seem more promising out-of-equilibrium 
systems, since they both take much longer to thermalize, but oscillations 
only become dominant so far from the wall as to be unobservable. Schemes 
to amplify the oscillating potential using a planar \cite{ellingsen09b} and 
cylindrical \cite{ellingsen10} cavity were explored, of which the latter showed 
some promise for observing the effect for Rydberg atoms. A related possibility 
seems to be the use of excited media \cite{sambale09b}. Casimir-Lifshitz 
and Casimir-Polder forces between bodies held at different temperatures, the
latter exhibiting regions of repulsion, were famously considered in
\cite{antezza1,antezza2,golyk}.
We will not consider 
these thermal non-equilibrium systems further herein.

The interaction of a classical dipole with a conducting screen containing an
aperture is a problem that can be exactly solved when the aperture is a
circle or an infinite slit, and the dipole is on the symmetry axis, and is
oriented perpendicular to the screen.  The circular problem
is essentially equivalent to the classical problem of the electrification of a
conducting disc. When the dipole is sufficiently close to the aperture,
it is repelled by the screen, which follows from the fact that the energy
must vanish at both infinity and zero, by symmetry.
 The solution is reviewed in \sref{sec:classical}.

Such solutions motivated the numerical discovery by Johnson's group 
\cite{Levin:2010zz}
that a sufficiently elongated conducting cylinder or ellipsoid is repelled 
by quantum vacuum
forces when the elongated body is on the symmetry axis and is 
sufficiently close to
the screen.  The classical symmetry argument no longer applies, however, 
since the energy need not vanish when the object is centered in the aperture.
This result strongly
suggests that there will be a repulsive Casimir-Polder force between a 
sufficiently anisotropic atom and a conducting screen with an aperture.  
 Analytically, one can demonstrate that such repulsion occurs between 
an anisotropic
atom and a dilute anisotropic screen with an aperture (\sref{sec:cpatoms}), 
and between such an atom
and a conducting half-plane or wedge (\sref{sec:cpwedge}), or at sufficiently 
large distance, a  cylinder (\sref{sec:cpcyl}).  
The same should occur between an atom and a screen possessing a slit, 
since the three-body
corrections to the half-plane result are expected to be small.

The outlook for further interesting examples of Casimir and Casimir-Polder
repulsion will be sketched in \sref{concl}, as well as the possibilities for
experimental verification.

\section{Repulsion due to intermediate material}
\label{sec:lifshitz}
There have been many derivations of the Lifshitz formula for the 
Casimir force between parallel, isotropic, dielectric slabs (of infinite
thickness), separated by a medium  possessing an electric
permittivity, for example, \cite{lifshitz61,schwinger78,miltonbook,mostbook,
miltonrev,lambrecht06,milton-brevik,Bordag:2012zz}.  Specifically, 
consider a dielectric function in the following form
\be
\varepsilon(\mathbf{r})=\left\{\begin{array}{cc}
\epsilon_1,&z<0,\\
\epsilon_3,&0<z<a,\\
\epsilon_2,&a<z.
\end{array}\right.\label{dielectricconst}
\ee
Dispersion is included in that all three permittivities can depend on
the (imaginary) frequency $\zeta=-\rmi\omega$.  
The energy and pressure on the slabs
are expressed in terms of the transverse electric (TE) 
reflection coefficients at the two interfaces,
\be 
r_{\rm TE}=\frac{\kappa_3-\kappa_1}{\kappa_3+\kappa_1},\quad
r'_{\rm TE}=\frac{\kappa_3-\kappa_2}{\kappa_3+\kappa_2},
\ee
and the transverse magnetic (TM) reflection coefficients,
which are obtained from these by replacing $\kappa_a\to
\kappa'_a=\kappa_a/\epsilon_a$.  Here
\be \kappa^2_a=k^2+\zeta^2\varepsilon_a(\rmi\zeta),\quad a=1,2,3, 
\ee
where $k$ is the magnitude of the transverse wavevector,
and we have made a Euclidean rotation, $\omega=\rmi\zeta$. 

The quantum fluctuation energy/area or Lifshitz energy/area for this
configuration is
\be
\fl \mathcal{E}=\frac1{4\pi^2}\int_{0}^\infty 
\rmd\zeta\int_0^\infty \rmd k\, k\left[
\ln\left(1-r_{\rm TE}r_{\rm TE}'\rme^{-2\kappa_3 a}\right)+
\ln\left(1-r_{\rm TM}r_{\rm TM}'\rme^{-2\kappa_3 a}\right)\right].
\label{lifshitz}
\ee
The force per area, or pressure, on plate 2 is 
\bea
\fl P=-\frac1{4\pi^2}\int_{0}^\infty\rmd \zeta \int_0^\infty 
\rmd k^2 \kappa_3
\left[\left((r_{\rm TE} r'_{TE})^{-1}\rme^{2\kappa_3 a}-1\right)^{-1}
+\left((r_{\rm TM} r'_{TM})^{-1}\rme^{2\kappa_3 a}-1\right)^{-1}\right].\nn\\
\eea

In the usual situation considered, 
the intervening material between the slabs
is rather dilute, so $\epsilon_1$, $\epsilon_2 >\epsilon_3$, and then
$r_{\rm TE}$, $r'_{\rm TE}$ are both negative, 
while $r_{\rm TM}$, $r'_{\rm TM}$ are both positive, 
and $(r_{\rm TE} r'_{\rm TE})^{-1}$, $(r_{\rm TM} r'_{\rm TM})^{-1}$
are both greater than one.  Thus the Lifshitz force is necessarily 
attractive. But if the intermediate material has an intermediate value of 
the permittivity, $\epsilon_1>\epsilon_3>\epsilon_2$, 
then both $(r_{\rm TE} r'_{\rm TE})^{-1}$, 
$(r_{\rm TM} r'_{\rm TM})^{-1}$ are negative, and the force is repulsive.
Actually in this case, the repulsion arises from the attraction of the 
intermediate
material (typically a fluid) into the intervening space.  This phenomenon is
well known, and remarked on in the earliest papers \cite{lifshitz61}.
Of course, the above conclusions hold only if the permittivities obey the
appropriate inequalities for all frequencies.  Real materials may possess
regions where the inverted behavior occurs, in which case repulsion only
can be achieved if that inverted inequality occurs over a sufficiently large frequency
range.

Recently, this repulsion was observed in an experiment by Munday, Capasso, and
Parsegian, who used bromobenzene between two gold surfaces, or between a silica
and a gold surface \cite{capasso09}.  Qualitative agreement with 
expectations was seen.
In particular, a repulsive force was observed between the gold and the silica.
There were a number of earlier experiments of this kind 
\cite{milling96,meurk97,lee01,lee02,feiler08}. But, in fact, the definitive 
experiment by Sabisky and Anderson which accurately verified the Lifshitz
theory \cite{sabisky} was exactly of this ``inverted'' type.
They determined the Casimir force between a substrate (CaF$_2$, SrF$_2$,
BaF$_2$), a liquid helium film, and helium vapor as a function of the thickness of the
film, determined by acoustic interferometry. (This geometry was already
considered in \cite{lifshitz61}.) The result 
was remarkably consistent with the Lifshitz
theory from 100 to 2500 nm.  (The earlier version of their results
\cite{anderson} was shown to be consistent with the theory of van der Waals
attraction by Richmond and Ninham \cite{richmond1,richmond2}, based on a code implementing
the Lifshitz formula by Ninham and Parsegian.)  Of course, the attractive
force between the substrate and the helium film, which is responsible for the
film climbing the interface, may be interpreted as a repulsion between the
substrate and the less dense helium gas.  Mention should also be made of
the experiment of Hauxwell and Ottewill \cite{hauxwell}, who measured the
thickness of alkanes on water.  Retardation effects in that case can lead
to repulsion \cite{ninham-parsegian}.  The melting of water ice exhibits
the same general phenomenon \cite{elbaum}.

A review of these Lifshitz repulsion phenomena, stressing the importance of
retardation in giving rise to the repulsion is given in \cite{bostrom2012}.
While the practical applicability of this type of repulsion in micro- and 
nanomechanical devices seems limited, 
some recent suggestions include repulsion between sheets of graphene upon 
introducing interspatial hydrogen gas \cite{bostrom2012a}.

\section{Repulsion in self-stress}\label{sec:boyer}
\begin{figure}
\centering
\epsfig{figure=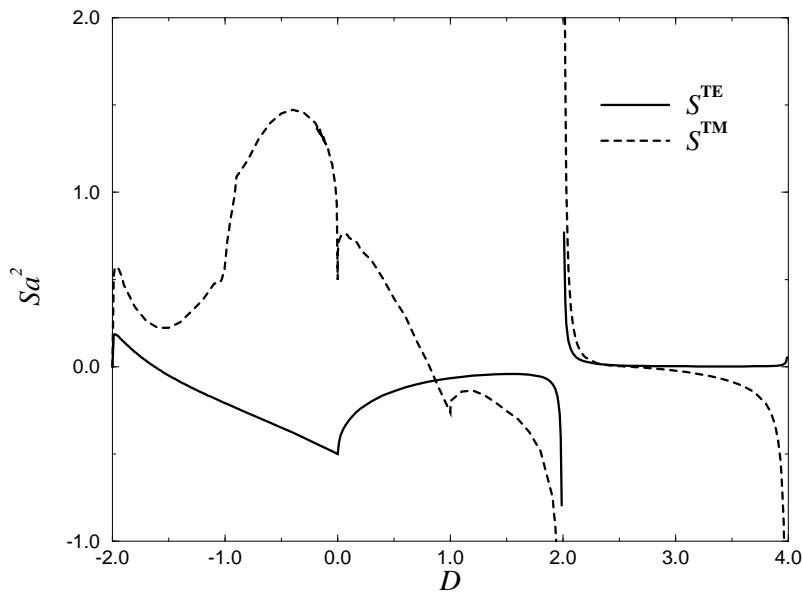,height=12cm,angle=270}
\caption{\label{dim} Dimensional dependence of the TE (Dirichlet) and
TM Casimir self stresses $S$ of perfect spherical shells of radius $a$.  These results were
obtained by dimensional continuation.  The electromagnetic case (really
only unambiguously defined in $D=3$ spatial dimensions) is obtained
by adding the TE and TM contributions and excluding the $l=0$ mode;
Boyer's result (\ref{boyer}) is reproduced.}
\end{figure}

Casimir speculated that quantum fluctuations would also cause a perfectly
conducting sphere to experience an ``attractive'' self-stress, that is,
the Casimir self-energy of such a spherical shell would be negative 
\cite{Casimir53}.  However, when Boyer carried out the impressive calculation
of the quantum self-energy of a perfectly conducting spherical shell, he found
a repulsive result \cite{boyer},
\be
E_{\rm sp}=+\frac1{a}0.04618\dots,\label{boyer}
\ee
where the last digits are the result of subsequent verifications of Boyer's
surprising result \cite{davies,balian,mds}.  The same repulsion is exhibited
for the self-energy of a Dirichlet sphere, where the fluctuations are in a
scalar field rather than in the electromagnetic field \cite{bender-milton,
  lesseduarte1,lesseduarte2,lesseduarte3}.  In fact there is a systematic
behavior for hyperspheres in any dimension \cite{bender-milton,miltond};
for the transverse electric (TE, or Dirichlet) modes, the energy is
positive (repulsive) for any spatial dimension $D$ in the interval
$2<D<4$, while the transverse magnetic (TM) mode changes sign at $D=2.6$,
being attractive for $2.6<D<4$. See \fref{dim}.
 There has been some dispute recently
about the calculability of scalar self energies for the
spherical geometry, because it is true
that in the electromagnetic case, certain divergences cancel, which is
not the case for say Dirichlet boundaries \cite{kolomeisky,fullingsp}.  
However,
in fact, the scalar cases are completely unambiguous; that is, the
divergent terms can be extracted in a universal Weyl expansion, as will
be discussed in a future publication \cite{miltonsp}.

\begin{figure}
\centering
\includegraphics{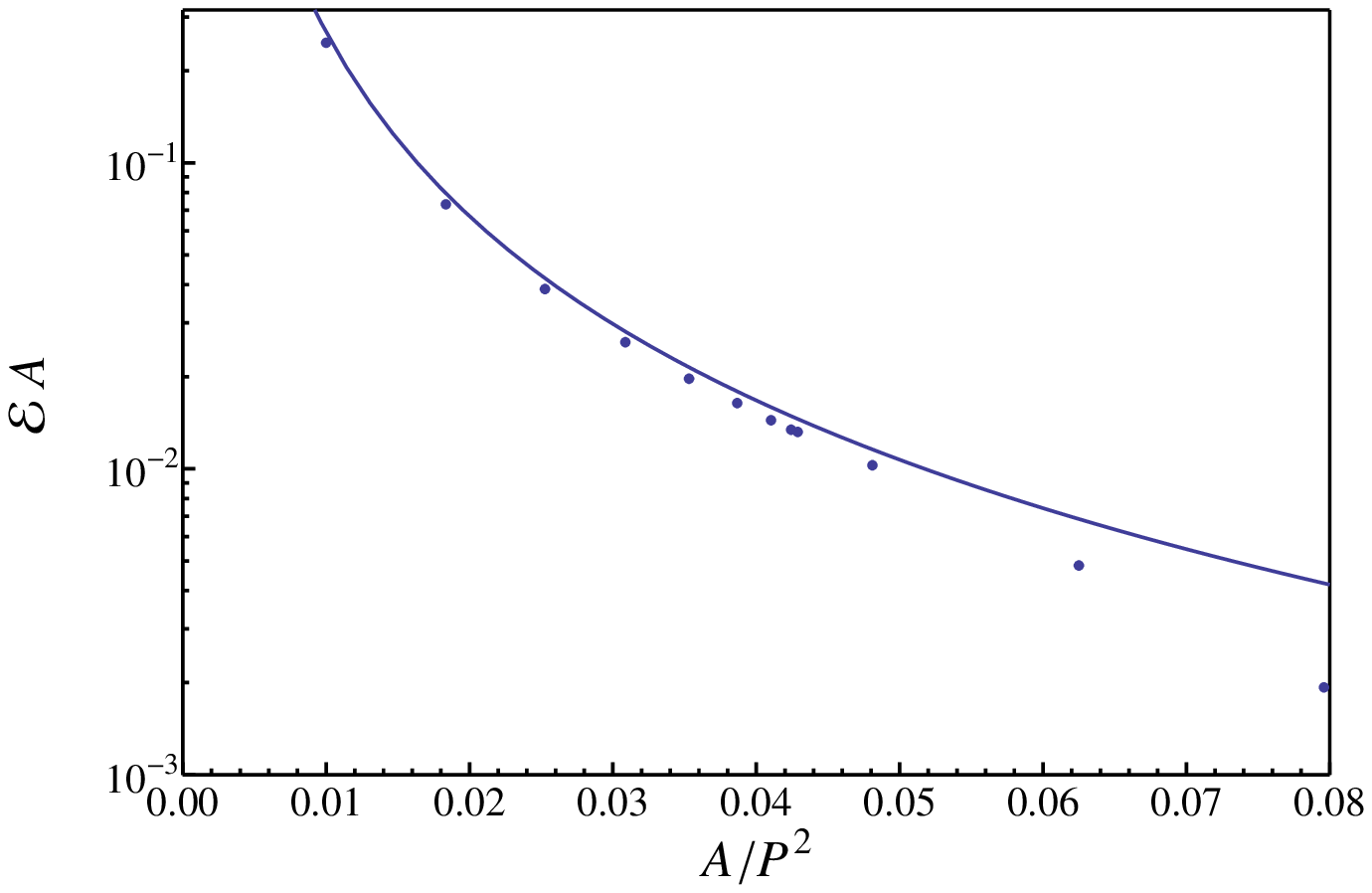}
\caption{\label{triangle} Scalar Casimir energies $\mathcal{E}$ per unit length,
scaled by the cross-sectional area $A$, for infinite cylinders having
triangular, square, and circular cross sections, obeying Dirichlet boundary
conditions on the surfaces.  The abscissa is the area $A$ divided by the
square of the perimeter $P$ of the cross section.
The Weyl volume, surface, and corner divergences
have been unambiguously subtracted.  Besides the circle and square,
the points represent equilateral, 
right-isosceles, and hemiquilateral triangular cross sections, as well as
right triangles calculated numerically.  The curve represents the PFA 
(proximity force approxiation) result.}
\end{figure}

Recently it has been recognized that finite and analytic Casimir 
energies could be
computed for infinite cylinders and finite prisms of square, equilateral,
right-isoceles, and bisected-equilateral cross sections \cite{abalo1}, as
well as for three ``integrable'' tetrahedra \cite{abalo2}. The self-energies
for infinite Dirichlet cylinders (including that of circular cross section
\cite{deraadcyl,gos},
as well as those of other triangles computed numerially) were all repulsive,
and lay on a universal curve \cite{abalo1}, as shown in \fref{triangle}.
  The Dirichlet energies for 
tetrahedra, cube \cite{lukosz1,lukosz2,lukosz3,ambjorn,ruggiero1,ruggiero2}, 
and finite (sufficiently short) prisms are negative (attractive).  Neumann 
boundary conditions generally give attractive results as well,
as do perfectly conducting electromagnetic boundary conditions for 
cylinders. These energies are all referring to
interior modes of the cavity only, except for spheres and cylinders,
where, to exclude curvature divergences, exterior modes are included as well. 
 The new systematics described in \cite{abalo1, abalo2} go
a long way for shedding light on the general behavior of Casimir 
self-energies, both in sign and magnitude,
which have remained elusive for decades.

Because there are no unremovable curvature divergences,
Casimir energies may be computed
unambiguously in these situations. However, the physical meaning of the results is unclear.
Certainly, if one surface is pulled away from a tetrahedral box, it will
experience an attractive force due to the closest elements of the surfaces.
The self-energy represents the energy required to assemble the complete structure,
but it is unclear how such a thing could ever be measured.

\section{Magnetic repulsion}\label{sec:magnetic}
Boyer also discovered \cite{boyer74} that a perfectly  electrically-conducting plate 
repels
a perfectly magnetically-conductive plate, that is, for one plate, the 
electric permittivity is taken to infinity, while for the second plate,
the magnetic permeability is taken to infinity.  
Indeed, it is straightforward
 to show that the Lifshitz energy per area between parallel
dielectric and diamagnetic  slabs, separated by a vacuum gap 
of thickness $a$, is
\be
\mathcal{E}_{\varepsilon\mu}=\frac1{8\pi^3}\int_0^\infty \rmd\zeta\int \rmd^2k\bigg[\ln\left(
1-r_1 r_2'\rme^{-2\kappa a}\right)
+\ln\left(1-r_1' r_2\rme^{-2\kappa a}\right)\bigg],\label{eem}\ee
where
\be
r_i=\frac{\kappa-\kappa_i}{\kappa+\kappa_i},\quad r_i'=
\frac{\kappa-\kappa'_i}{\kappa+\kappa'_i},\ee
with
\be
\fl \kappa^2=k^2+\zeta^2,\quad 
\kappa^2_1=k^2+\varepsilon\zeta^2, \quad \kappa_1'
=\kappa_1/\varepsilon,\quad \kappa^2_2=k^2+\mu\zeta^2,\quad\kappa_2'
=\kappa_2/\mu.
\ee
This means in the perfect reflecting limit,
$\varepsilon\to\infty$, $\mu\to\infty$,
\be
\mathcal{E}_{\rm Boyer}=+\frac78\frac{\pi^2}{720 a^3},
\ee
 we get Boyer's repulsive result \cite{boyer74}, $-7/8$ times
the Casimir attraction that follows from the appropriate limit of
\eref{lifshitz}.

However, materials with such a large magnetic response over a sufficiently
large frequency range are difficult to manufacture.  One might think that
artificial materials, metamaterials, would be a route to magnetic
repulsion \cite{Henkel,Pirozhenko,Rosa,Rosa2,Zhao,comment,commenta}.
Despite some early optimism, the conclusion seems to be settled that 
repulsion is impossible between metamaterials made from dielectric and 
metallic components \cite{Yannopapas, Silveirinha,McCauley}.  For
recent attempts using dielectric/magnetic setups see 
\cite{maslovski,zeng,grushin}, who consider nanowires, ferrites,
and topological insulators, respectively.

\section{Classical repulsion}
\label{sec:classical}
The discussion in this section is based on \cite{rep1}.
\subsection{Repulsion between dipoles}
\begin{figure}
 \begin{center}
\includegraphics[scale=1]{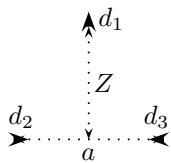}
\caption{\label{fig3} Configuration of three dipoles, two of which are
antiparallel, and one perpendicular to the other two.}
\end{center}
\end{figure}

Of course, repulsion occurs in classical situations.  Not only do like
charges repel, but electric dipoles exhibit regimes in which repulsion
occurs as well. Consider the situation illustrated in \fref{fig3}.
Here we have two dipoles, of strength $d_2$ and $d_3$ lying along the $x$
axis, separated by a distance $a$.  A third dipole of strength $d_1$ lies
along the $z$ axis, a distance $Z$ from the line of the dipoles 2 and 3.  
If the two parallel dipoles are oppositely directed and of equal strength,
\be
\mathbf{d}_2=-\mathbf{d}_3=d_2 \mathbf{\hat x}, 
\ee
and are equally distant from the $z$ axis,
and the dipole on the $z$ axis is directed along that axis,
\be
\mathbf{d}_1=d_1\mathbf{\hat z},
\ee
the force on that dipole is along the $z$ axis:
\be
F_z=3ad_1d_2\frac{a^2/4-4 Z^2}{(Z^2+a^2/4)^{7/2}},
\ee
which changes sign at $Z=a/4$.  That is, for distances $Z$ larger than this,
the force is attractive (in the $-z$ direction) while for shorter distances
the force is repulsive (in the $+z$ direction).  Evidently, by symmetry,
the dipole-dipole energy vanishes at $z=0$.  Consistent with Earnshaw's
theorem \cite{stratton}, the point where the force vanishes is an unstable point with
respect to deviations in the $x$ direction.

\subsection{Three dimensional aperture interacting with dipole}\label{sec4b}
More interesting is the interaction of a dipole with a conducting screen containing
an aperture, for example, a slot or a circular hole.  For the case of a
dipole, polarized along the symmetry
axis, a distance $Z$ directly above a circular aperture of radius
$a$  in a conducting plate, the problem can be solved in closed form. 

The free three-dimensional
 Green's function in cylindrical coordinates has the representation
\be
G_0(\mathbf{r,r'})=
\frac1{\sqrt{\rho^2+(z-z')^2}}=\int_0^\infty \rmd k\,J_0(k\rho)
\rme^{-k|z-z'|},\label{freegcc}
\ee
from which we can immediately find the Green's function,
\be
G(\mathbf{r,r'})=G_0(\rho,z-z')-G_0(\rho,z+z'),
\quad \rho=\sqrt{(x-x')^2+(y-y')^2},
\ee so constructed that
\be G(\rho,z=0,z')=0.
\ee
Using this, 
we can calculate the electrostatic potential at any point above the $z=0$ plane
to be
\be
\phi(\mathbf{r})=\int_{z>0}(\rmd\mathbf{r'})\,G(\mathbf{r,r'})\rho(\mathbf{r'})
+\frac1{4\pi}\int_{\rm ap}\rmd S'\frac\partial{\partial z'}G(\mathbf{r,r'})\bigg|_{z'=0}
\phi(\mathbf{r'}),
\ee
where the volume integral is over the charge density of the dipole,
\be
\rho(\mathbf{r})=-\mathbf{d}\cdot\bnabla\delta(\mathbf{r-R}),\quad \mathbf{R}=(0,Z).
\ee
The surface integral extends only over the aperture because the potential
vanishes on the conducting sheet. Using the addition theorem for Bessel
functions, we then find for the potential above
the plate
\bea
\phi(\mathbf{r_\perp},z>0)&=&d\left[\frac{z-Z}{[r_\perp^2+(z-Z)^2]^{3/2}}+
\frac{z+Z}{[r_\perp^2+(z+Z)^2]^{3/2}}\right]\nn\\
&&\quad\mbox{}+\int_0^\infty \rmd k\,k\, \rme^{-kz}J_0(kr_\perp)\Phi(k),
\eea
where the Bessel transform of the potential in the aperture is
\be
\Phi(k)=\int_0^\infty \rmd\rho\,\rho\, J_0(k\rho)\phi(\rho,0).
\ee
Below the aperture there is no contribution from the dipole
if we similarly integrate below the $z=0$ plane,
\be
\phi(\mathbf{r_\perp},z<0)=
\int_0^\infty \rmd k\,k\, \rme^{kz}J_0(kr_\perp)\Phi(k).
\ee
Then we obtain two  integral equations resulting from the continuity of the 
$z$-component
of the electric field in the aperture and the vanishing of the potential on
the conductor:
\numparts
\bea
d\frac{r_\perp^2-2Z^2}{[r_\perp^2+Z^2]^{5/2}}=\int_0^\infty \rmd k \,k^2 J_0(k
r_\perp)\Phi(k),\quad r_\perp<a,\\
0=\int_0^\infty \rmd k\,k J_0(kr_\perp)\Phi(k),\quad r_\perp>a.
\eea
\endnumparts

The solution to these equations is given in Titchmarsh's book \cite{titchmarsh}, 
and after a bit of manipulation we obtain
\be
\Phi(k)=-\left(\frac{2}{\pi ka}\right)^{1/2}d\int_0^1 \rmd x\,x^{3/2}
J_{1/2}(xka)\frac{2Z/a}{(x^2+Z^2/a^2)^2}.
\ee
From this, we can work out the energy of the system from
\be
U=-\frac{d}2E_z(0,Z)=\frac{d}2\frac{\partial\phi}{\partial z}\bigg|_{z=Z,x=0},
\label{energy}
\ee
where the factor of 1/2 comes from the fact that this must be the energy
required to assemble the system.  In computing this energy we must, of course,
drop the self-energy of the dipole due to its own field.  We 
encounter the integral
\be
\int_0^\infty \rmd k\,k^{3/2}\rme^{-kZ}J_{1/2}(kax)=2\sqrt{\frac{2xa}\pi}
\frac{Z}{(x^2a^2+Z^2)^2},
\ee
and then we can express the energy in closed form:
\be
U=-\frac{d^2}{8Z^3}+\frac{d^2}{4\pi Z^3}\left[\arctan\frac{a}Z
+\frac{Z}a\frac{1+8/3(Z/a)^2-(Z/a)^4}{(1+Z^2/a^2)^3}\right].
\ee
This is always negative, but vanishes at infinity and at zero:
\be
Z\to 0:\quad U\to-\frac4{5\pi}d^2\frac{Z^2}{a^5}.
\ee
This means that for some value of $Z\sim a$ the force changes from
attractive to repulsive.  We find that the force changes
sign at $Z=0.742358a$. 

The reason why the energy vanishes when the dipole is centered in the
aperture is clear: Then the electric field lines are perpendicular to the
conducting sheet on the surface, and the sheet could be removed without
changing the field configuration.

A similar calculation, with qualitatively identical results, occurs
for two conducting half planes separated by an infinite slit \cite{rep1}.

\section{Casimir-Polder repulsion between atoms}
\label{sec:cpatoms}
\begin{figure}
\begin{center}
\includegraphics{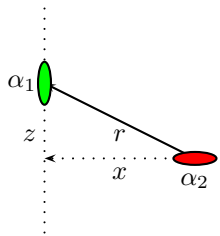}
\caption{\label{pol-atoms} Casimir-Polder interaction between two atoms
of polarizability $\balpha_1$ and $\balpha_2$ separated by a distance $r$.
Atom 1 is predominantly polarizable in the $z$ direction, while atom 2
is predominantly polarizable in the $x$ direction.  The force on atom 1
in the $z$ direction becomes repulsive sufficiently close to the polarization
axis of atom 2 provided both atoms are sufficiently anisotropic.}
\end{center}
\end{figure}

We now turn to the quantum repulsion.  This was first revealed by the
numerical results of Levin \etal \cite{Levin:2010zz}, followed by
some analytical work \cite{maghrebi}.  Our group is in the midst
of an extensive analysis of Casimir-Polder repulsion.
The following discussion first appeared in \cite{rep2}.
The interaction between two polarizable atoms, described by general
polarizabilities $\balpha_{1,2}$, with the relative separation
vector given by  $\mathbf{r}$ is \cite{CP1,CP2}
\be
\fl U_{\rm CP}=-\frac1{4\pi r^7}\left[\frac{13}2\Tr\balpha_1\cdot
\balpha_2-28\Tr(\balpha_1\cdot\mathbf{\hat r})(\balpha_2
\cdot\mathbf{\hat r})+\frac{63}2(\mathbf{\hat r}\cdot\balpha_1
\cdot\mathbf{\hat r})(\mathbf{\hat r}\cdot\balpha_2
\cdot\mathbf{\hat r})\right].\label{generalcp}
\ee
This formula is easily rederived by the multiple scattering
technique as explained in \cite{brevikfest}.
This reduces, in the isotropic case, $\balpha_i=\alpha_i\boldsymbol{1}$,
to the usual Casimir-Polder (CP) energy, 
$U_{\rm CP}=-\frac{23}{4\pi r^7}\alpha_1\alpha_2$.
Suppose the two atoms are only polarizable in perpendicular directions,
$\balpha_1=\alpha_1\mathbf{\hat z \hat z}$,
$\balpha_2=\alpha_2\mathbf{\hat x \hat x}$, as shown in \fref{pol-atoms}.
 Choose atom 2 to be at the origin. 
Then, in terms of the polar angle $\cos\theta=z/r$, the $z$-component of
the force on atom 1 is
\be
F_z=-\frac{63}{8\pi}\frac{\alpha_1\alpha_2}{x^8}\sin^{10}\theta\cos\theta
(9-11\sin^2\theta).
\ee
Here, we are considering motion for fixed $x=r\sin\theta$,
in the $y=0$ plane.
Evidently, the force is attractive at large distances, vanishing
as $\theta\to0$, but it must change sign at small values of $z$ for fixed $x$,
since the energy also vanishes as $\theta\to\pi/2$.  The force component
in the $z$ direction vanishes when $\sin\theta=3/\sqrt{11}$ or $\theta=1.130$
or 25$^\circ$ from the $x$ axis.

No repulsion occurs if one of the atoms is isotropically polarizable.
If both have cylindrically symmetric anisotropies, but with respect
to perpendicular axes,
\be
\balpha_1=(1-\gamma_1)\alpha_1\mathbf{\hat z \hat z}+\gamma_1\alpha_1\boldsymbol{1},\quad 
\balpha_2=(1-\gamma_2)\alpha_2\mathbf{\hat x \hat x}+\gamma_2\alpha_2\boldsymbol{1},
\ee
it is easy to check that, if both are sufficiently anisotropic, repulsion will occur.
For example, if $\gamma_1=\gamma_2$ repulsion in the $z$ direction will take place close to the plane
$z=0$ if $\gamma\le0.26$.

After \cite{rep2} was submitted, a paper by  Shajesh and Schaden
\cite{SS} appeared, which rederived these results, and then
went on to extend the calculation to Casimir-Polder repulsion by an anisotropic
dilute dielectric sheet with a circular aperture.  The authors quite correctly
point out that the statement in \cite{rep1}  that no repulsion is possible in
the weak-coupling regime is erroneous.  That inference was based on isotropic
media; anisotropy is necessary for repulsion.

\begin{figure}
 \begin{center}
\includegraphics[scale=1]{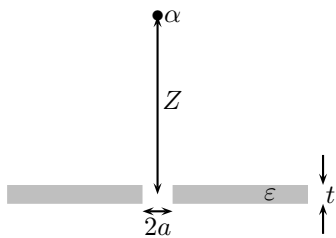}
\caption{\label{fig2} Three-dimensional geometry of a polarizable
atom a distance $Z$ above a dielectric slab of thickness $t$ with a circular
aperture of radius $a$.}
\end{center}
\end{figure}

Here we extend the calculations of \cite{SS}.  We consider an anisotropic
polarizable atom directly above a tenuous anisotropic slab containing a
circular aperture, as shown in \fref{fig2}.
Here we assume that the atom is only polarizable in the $z$ direction,
\be
\balpha_1=\alpha_1\mathbf{\hat z\hat z},
\ee
while the slab is composed of atoms only transversely polarizable,
\be
\balpha_2=\alpha_2(\mathbf{\hat x\hat x+\hat y\hat y}).
\ee
Starting from (\ref{generalcp}), we use cylindrical coordinates with
origin at the center of the aperture, and find the interaction energy between
atom 1 at $(0, 0, Z)$, and atom 2 at $(\rho, 0, z)$
to be
\be
U_{\rm CP}=-\frac{63\alpha_1\alpha_2}{8\pi r^7}\left(\frac{Z-z}r\right)^2
\left(\frac\rho{r}\right)^2,
\ee
which when integrated over the slab made up of the type-2 atoms
gives the quantum interaction energy
\bea
E&=&-\frac{63\alpha_1\alpha_2 n_2}{8\pi}\int_a^\infty \rmd \rho\,\rho
\int_0^{2\pi}\rmd \theta\int_{-t/2}^{t/2}\rmd z\frac{(Z-z)^2\rho^2}
{\left[(Z-z)^2+\rho^2\right]^{11/2}}\nn\\
&=&-\frac{63\alpha_1\alpha_2 n_2}{2a^4}\left[e(Z/a+t/2a)-e(Z/a-t/2a)
\right],
\eea
where $n_2$ is the number density of atoms in the slab, $t$ is the
thickness of the slab, and the atom is located a distance $Z$ directly
above the aperture.  Here we have defined the function
\be
e(x)=\frac{x^3}5\frac{15+14 x^2+4 x^4}{(1+x^2)^{7/2}}.
\ee

Such an atom sufficiently close to the aperture experiences a
repulsive force.  Define a dimensionless parameter $\delta$ that measures the
height of the atom above the top of the aperture,
\be
Z=\frac{t}2+a\delta.
\ee
For a thick aperture, $t/a\gg1$, it is easy to check that that the
force changes sign very close to the opening of the aperture,
\be
\delta=\frac{\sqrt{2}}3\left(\frac{t}a\right)^{-5/2};
\ee 
for example, when $t/a=10$, $\delta =1.5\times 10^{-3}$.  When the
aperture is very thin, $t/a\ll1$, the value of $\delta$ for which 
repulsion sets in becomes independent of $t/a$, $\delta=0.5566$,
which again agrees with the result of \cite{SS}.

\section{Casimir-Polder force between an atom and a conducting wedge}
\label{sec:cpwedge}

\subsection{General formula for Casimir-Polder interaction}
\label{sec5}
Now we turn to the CP interaction between an atom
and a dielectric or conducting body.  Our starting point is the general
expression for the vacuum energy \cite{brevikfest}
\be
U=\frac{i}2\Tr\ln{\bGamma\bGamma}_0^{-1},
\ee
where $\bGamma$ is the full Green's dyadic for the problem, and
$\bGamma_0^{-1}$ is the inverse of the free Green's dyadic, namely
\be
\bGamma_0^{-1}=\frac1{\omega^2}\bnabla\times\bnabla\times-\boldsymbol{1}.
\ee
In the presence of a potential $\mathbf{V}$, the full Green's dyadic has
the symbolic form
\be
\bGamma=(\boldsymbol{1}-\bGamma_0\mathbf{V})^{-1}\bGamma_0.
\label{Gammaexp}
\ee

Here we are thinking of the interaction between a dielectric medium,
characterized by an isotropic permittivity, so $V_1=\varepsilon-1$, 
and a polarizable
atom, represented by a polarizability dyadic, as shown in \fref{fig2},
\be
\mathbf{V}_2=4\pi\balpha\delta(\mathbf{r-R}),
\ee
where $\mathbf{R}$ is the position of the dipole.  We are only interested
in a single interaction with the latter potential, so we have for the interaction
energy
\be
U_{12}=\Tr \mathbf{V}_2\frac\delta{\delta V_1}
\left[-\frac{i}2\ln\left(1-\bGamma_0V_1
\right)\right]
=\frac{i}2\Tr\left(\bGamma_1-\bGamma_0\right)\mathbf{V}_2,
\ee
where we have used (\ref{Gammaexp}) for the potential $V_1$ describing
 the dielectric slab plus aperture and we have
subtracted the term that represents the self-energy of the atom
with its own field.
This subtraction happens automatically if we start from the ``$TGTG$'' form,
\be
\fl U_{12}=-\frac{i}2\Tr\ln\left(\boldsymbol{1}
-\bGamma_1\mathbf{V}_1\bGamma_2\mathbf{V}_2\right)
\approx\frac{i}2\Tr \bGamma_1 \mathbf{V}_1\bGamma_0\mathbf{V}_2
=\frac{i}2\Tr \left(\bGamma_1-\bGamma_0\right)\mathbf{V}_2,
\ee
because $\mathbf{V}_2$ is weak.
This implies the Casimir-Polder expression for the interaction between
the polarizable atom and the dielectric
\be
U_{\rm CP}=-\int_{-\infty}^\infty \rmd\zeta \tr\balpha\cdot(\bGamma
-\bGamma_0)(Z,Z).\label{alphagamma}
\ee
(Note that although we use Gaussian units otherwise, the Green's dyadics
are still expressed in terms of Heaviside-Lorentz units; otherwise, factors
of $4\pi$ appear.)

\subsection{Wedge calculation}

\begin{figure}
 \begin{centering}
\includegraphics[scale=1]{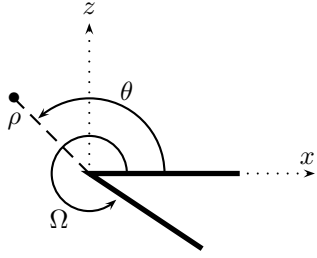}
\caption{\label{figwedge}  Polarizable atom, located at polar coordinates  
$\rho$, $\theta$, within a conducting wedge with dihedral angle $\Omega$.}
 \end{centering}
\end{figure}

The interaction between a polarizable atom and a perfectly conducting
half-plane is a special case of the vacuum interaction between such an atom
and a conducting wedge.  For the case of an isotropic atom, this was
considered by Brevik, Lygren, and Marachevsky \cite{blm}.  (This followed
on earlier work by Brevik and Lygren \cite{bl} and DeRaad and Milton
\cite{deraadcyl}.)
In terms of the opening dihedral angle of the wedge $\Omega$,
which we describe
in terms of the variable $p=\pi/\Omega$, the electromagnetic Green's
dyadic has the form (here the translational direction is denoted by $y$,
and one plane of the wedge lies in the $z=0$ plane, the other intersecting
the $xz$ plane on the line $\theta=\Omega$---see \fref{figwedge})
\bea
\fl\bGamma(\mathbf{r,r'})=2p\sum_{m=0}^\infty{}'\int\frac{\rmd k}{2\pi}
\bigg[-\boldsymbol{\mathcal{M}}\boldsymbol{\mathcal{M}}^{\prime*}
(\nabla_\perp^2-k^2)\frac1{\omega^2}
F_{mp}(\rho,\rho')\frac{\cos mp\theta \cos mp\theta'}\pi \rme^{\rmi k(y-y')} 
 \nn\\
\mbox{}+\boldsymbol{\mathcal{N}}\boldsymbol{\mathcal{N}}^{\prime*}
\frac1{\omega}
G_{mp}(\rho,\rho')\frac{\sin mp\theta \sin mp\theta'}\pi \rme^{\rmi k(y-y')}
\bigg].
\eea
The first term here refers to TE (H) modes, the second to TM (E) modes.
The prime on the summation sign means that the $m=0$ term is counted with
half weight.  In the polar coordinates in the $xz$ plane, $\rho$ and 
$\theta$, the H and E mode operators are
\numparts\label{hemodeops}
\bea
\boldsymbol{\mathcal{M}}=\boldsymbol{\hat \rho}\frac\partial{\rho\partial\theta}
-\boldsymbol{\hat\theta}\frac\partial{\partial \rho},\\
\boldsymbol{\mathcal{N}}=\rmi k\left(\boldsymbol{\hat \rho}
\frac\partial{\partial\rho}
+\boldsymbol{\hat\theta}\frac\partial{\rho\partial \theta}\right)
-\mathbf{\hat y}\nabla_\perp^2,
\eea
\endnumparts
where the transverse Laplacian is
\bea
\nabla_\perp^2=
\frac1\rho\frac\partial{\partial \rho}\rho\frac\partial{\partial\rho}
+\frac1{\rho^2}
\frac{\partial^2}{\partial \theta^2}.
\eea
In this situation, the boundaries are entirely in planes of constant 
$\theta$,
so the radial Green's functions are equal to the free Green's function
\be
\frac1{\omega^2}F_{mp}(\rho,\rho')=\frac1\omega G_{mp}(\rho,\rho')
=-\frac{\rmi\pi}{2\lambda^2}
J_{mp}(\lambda \rho_<)H^{(1)}_{mp}(\lambda \rho_>),
\ee
with $\lambda^2=\omega^2-k^2$.
We will immediately make the Euclidean rotation, $\omega\to \rmi\zeta$, where
$\lambda\to \rmi\kappa$, $\kappa^2=\zeta^2+k^2$, so the free Green's 
functions become $-\kappa^{-2}I_{mp}(\kappa\rho_<)K_{mp}(\kappa \rho_>)$.

We start by considering the most favorable case for CP repulsion, where the
atom is only polarizable in the $z$ direction, that is, only $\alpha_{zz}\ne0$.
In the static limit, the only component of the Green's dyadic that
contributes is
\bea
\fl\int\frac{\rmd\zeta}{2\pi}\Gamma_{zz}
=\frac{2p}{4\pi^3}\int \rmd k\,\rmd\zeta
\bigg\{\left[\zeta^2\sin^2\theta\sin^2mp\theta-k^2\cos^2\theta\cos^2mp\theta
\right]\nn\\
\quad\times\frac{m^2p^2}{\kappa^2\rho_<\rho_>}I_{mp}(\kappa\rho_<)K_{mp}(\kappa
\rho_>)\nn\\
\quad\mbox{}-\left[k^2\sin^2\theta\sin^2 mp\theta-\zeta^2\cos^2\theta
\cos^2mp\theta\right]I'_{mp}(\kappa\rho_<)K'_{mp}(\kappa \rho_>)\bigg\}.
\eea
Here we note that the off diagonal $\rho$-$\theta$ terms in $\bGamma$ cancel.
We have regulated the result by point-splitting in the radial coordinate.
At the end of the calculation, the limit $\rho_<\to\rho_>=\rho$ is to be
taken.

Now the integral over the Bessel functions is given by
\be
\int_0^\infty \rmd\kappa\,\kappa\, I_\nu(\kappa\rho_<)K_\nu(\kappa\rho_>)
=\frac{\xi^\nu}{\rho_>^2(1-\xi^2)},
\ee
where $\xi=\rho_</\rho_>$. After that the $m$ sum is easily carried out
by summing a geometrical series.  Care must also be taken with the $m=0$ term
in the cosine series.  The result of a straightforward calculation leads to
\be
\label{vacuum}
\int\frac{\rmd\zeta}{2\pi}\Gamma_{zz}=-\frac{\cos 2\theta}{\pi^2\rho^4}
\frac1{(\xi-1)^4}+\mbox{finite},
\ee
where the term divergent as $\xi\to1$
 may, through a similar calculation, be shown to
be that corresponding to the vacuum in absence of the wedge, that is, that
obtained from the free Green's dyadic.  Therefore, we must subtract this term
off, to obtain the static Casimir energy (\ref{alphagamma}),
which for this situation is
\bea
\fl U^{zz}_{\rm CP}=-\frac{\alpha_{zz}(0)}{8\pi}\frac1{\rho^4\sin^4p\theta}
\left[p^4-\frac{2}3p^2(p^2-1)\sin^2p\theta+\frac{(p^2-1)(p^2+11)}{45}\sin^4p\theta
\cos2\theta\right].\nn\\ \label{ucpzz}
\eea
This result can also be  derived from the closed form for the Green's function
given by Lukosz \cite{lukosz3}.

A small check of this result is that as $\theta\to 0$ (or $\theta\to\Omega$)
we recover the expected Casimir-Polder result for an atom above an infinite
plane:
\be
U_{\rm CP}^{zz}\to -\frac{\alpha_{zz}(0)}{8\pi Z^4},
\ee
in terms of the distance of the atom above the plane, $Z=\rho\theta$.
This limit is also obtained when $p\to1$, for when $\Omega=\pi$ we are
describing a perfectly conducting infinite plane.

A very similar calculation gives the result for
an isotropic atom, $\balpha=\alpha\boldsymbol{1}$, which was first given in
\cite{blm}:
\be
\fl U_{\rm CP}=-\frac{3\alpha(0)}{8\pi\rho^4\sin^4p\theta}\left[p^4-\frac23p^2(p^2-1)
\sin^2p\theta-\frac13\frac1{45}(p^2-1)(p^2+11)\sin^4 p\theta\right].
\ee
Note that this is not three times $U_{\rm CP}^{zz}$ in (\ref{ucpzz})
 because the $\cos 2\theta$ factor in the last term in the latter is replaced
by $-1/3$ here.  This case was reconsidered recently, for example, in
\cite{mendez}.
\begin{figure}
 \begin{centering}
\includegraphics[scale=1]{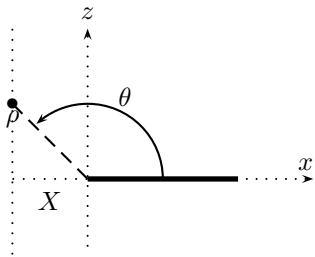}
\caption{\label{fighp}  Polarizable atom, above a half conducting plane,
free to move on a line perpendicular to the plane but a distance $X$ to
the left of the plane.}
 \end{centering}
\end{figure}

\subsection{Repulsion by a conducting half-plane}
\label{sec:cphp}
Let us consider the special case $p=1/2$, that is $\Omega=2\pi$, the case of
a semi-infinite conducting plane. This was the situation considered,
for anisotropic atoms, in recent papers  by Eberlein and Zietal
\cite{ce1,ce2}.  Note that in such a case, for the completely
anisotropic atom, $U_{\rm CP}^{zz}=0$ at $\theta=\pi$, which is obvious by symmetry.

\begin{figure}
 \begin{center}
\includegraphics[scale=1]{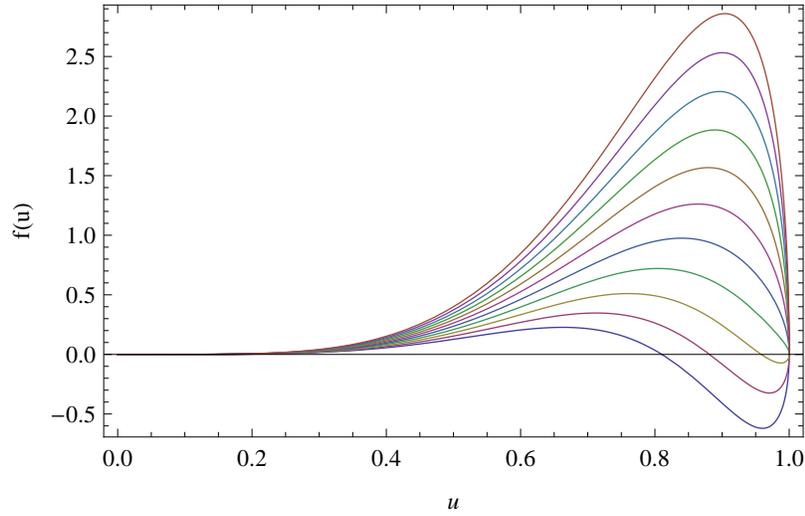}
\caption{\label{repsemi}                                           
The $z$-component of the force between an anisotropic
atom (with ratio of transverse to longitudinal polarizabilities $\gamma$)
and a semi-infinite perfectly conducting plane, $z=0$, $x>0$.
$F_z=-\alpha_{zz} f(u)/(32\pi X^5)$ in terms of the variable  
$u=X/\rho=-\cos\theta$. The atom lies on the line $y=0$, $x=-X$, 
and $\rho$ is the distance from the
edge of the plane to the atom.  Here, $f>0$ corresponds to an attractive
force on the $z$ direction, and $f<0$ corresponds to a repulsive force.
The different curves correspond to different values of $\gamma$, $\gamma=0$ to
1 by steps of 0.1, from bottom to top.
For $\gamma<1/4$ a repulsive regime always occurs when the atom is sufficiently
close to the plane of the conductor.}
\end{center}
\end{figure}

\begin{figure}
\begin{center}
\includegraphics[scale=1]{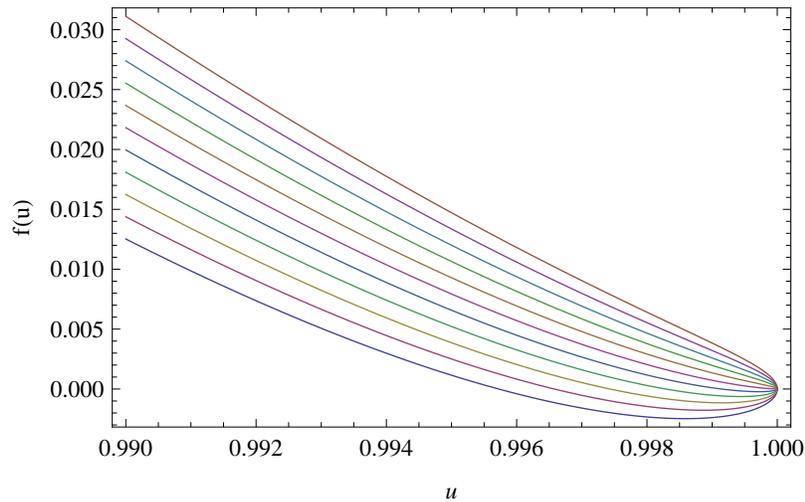}
\caption{\label{repsemi2}  Same as \fref{repsemi}.  The region 
close to the plane, $1\ge u\ge 0.99$,
 with $\gamma$ near the critical value of 1/4.
Here from bottom to top are shown the results for values of $\gamma$ from
0.245 to 0.255 by steps of 0.001.}
\end{center}
\end{figure}

 Consider a particle free to move along a
line parallel to the $z$ axis, a distance $X$ to the left of the
semi-infinite plane.  See \fref{fighp}.  The half-plane $z=0$, $x<0$
constitutes an aperture of infinite width. With $X$ fixed, we
can describe the trajectory by $u=X/\rho=-\cos\theta$,
which ranges from zero to one.  The polar angle is given by
\be
\sin^2\frac\theta2=\frac{1+u}2.
\ee
 The energy for an isotropic atom is given by
\be
U_{\rm CP}=-\frac{\alpha(0)}{32\pi}\frac1{X^4}V(u),
\ee
where
\be
V(u)=3u^4\left[\frac1{(1+u)^2}+\frac1{u+1}+\frac14\right].\label{uiso}
\ee
The energy for the completely anisotropic atom is
\be
V_{zz}=\frac13V(u)+\frac{u^4}2(1-3u^2).
\ee
Let us consider instead a cylindrically symmetric polarizable
atom in which
\be
\balpha=\alpha_{zz}\mathbf{\hat z\hat z}+\gamma\alpha_{zz}(\mathbf{\hat x
\hat x+\hat y\hat y})=\alpha_{zz}(1-\gamma)\mathbf{\hat z\hat z}
+\gamma\alpha_{zz}\boldsymbol{1},\label{gammapol}
\ee
where $\gamma$ is the ratio of the transverse polarizability to the
longitudinal polarizability of the atom.
Then the effective potential is
\be
(1-\gamma)V_{zz}+\gamma V,\label{effpot}
\ee
and the $z$-component of the force on the atom is
\be
\fl F^\gamma_z=-\frac{\alpha_{zz}(0)}{32\pi}\frac1{X^5}u^2\sqrt{1-u^2}
\frac{\rmd}{\rmd u}
\left[\frac12u^4(1-\gamma)(1-3u^2)+\frac13(1+2\gamma)V(u)\right],\label{fz}
\ee
where $V$ is given by (\ref{uiso}).  Note that the energy (\ref{effpot}), or the
quantity in square brackets in (\ref{fz}), only vanishes at $u=1$ ($\theta=\pi$, the
plane of the conductor) when $\gamma=0$. Thus, the symmetry argument given in
\cite{Levin:2010zz} applies only for the completely anisotropic case.
The force is plotted in figures \ref{repsemi}, \ref{repsemi2}.
It will be seen that if $\gamma$
is sufficiently small, when the atom is sufficiently close to
the plane of  the plate the
$z$-component of the force is repulsive rather than attractive.  The critical
value of $\gamma$ is $\gamma_c=1/4$.  This is a completely analytic exact
analog of the numerical calculations shown in \cite{Levin:2010zz},
where the interaction was considered
between a conducting plane with an aperture (circular
hole or slit), and a conducting cylindrical or ellipsoidal object.
Our calculation demonstrates that three-body effects are not required to
exhibit Casimir-Polder repulsion.  In fact, three-body effects are rather
small \cite{maghrebi}.

It is interesting to observe that the same critical value of $\gamma$
occurs in the nonretarded regime for a circular aperture, as follows from
a simple computation based on the result of \cite{ce3}.
For example, applying the result there for an atom with polarizability
given by (\ref{gammapol}) placed a distance $Z$ along the symmetry axis 
of a circular aperture of radius $a$ in a conducting plane gives an energy
\bea
\fl U=-\frac1{16\pi^2}\int_{-\infty}^\infty \rmd\zeta\,\alpha_{zz}(\zeta)
\frac1{Z^3}\Bigg\{(1+\gamma)\left(\frac\pi2+\arctan\frac{Z^2-a^2}{2aZ}\right)
\nn\\
\mbox{}+\frac{2aZ}{(Z^2+a^2)^3}\left[(1+\gamma)(Z^4-a^4)
-\frac83(1-\gamma)a^2Z^2 \right]\Bigg\}.
\eea
It is easy to see that this has a minimum for $z>0$, and 
hence there is a repulsive
force close to the aperture, provided $\gamma<\gamma_c=1/4$.

\subsection{Repulsion by a wedge}
\begin{figure}
\begin{center}
\includegraphics[scale=1]{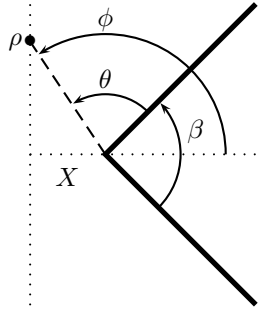}
\caption{\label{cpwedge}  A polarizable atom outside a perfectly   
conducting wedge of interior angle $\beta$.
The atom is located at polar angles $\rho$,
$\phi$ relative to the symmetry plane of the wedge.}
\end{center}
\end{figure}

It is very easy to generalize the above result for a wedge, $p>1/2$.  That 
is,
we want to consider a strongly anisotropic atom, with only $\alpha_{zz}$
significant, to the left of a wedge of opening angle
\be
\beta=2\pi-\Omega,
\ee
as shown in \fref{cpwedge}.
We want the $z$ axis to be perpendicular to the symmetry plane of the wedge
so the relation between the polar angle of the atom and the
angle to the symmetry line
is
\be
\phi=\theta+\beta/2,
\ee
where, as before, $\theta$ is the angle relative to the top surface of the
wedge.  Then, it is obvious that the formula for the Casimir-Polder energy
(\ref{ucpzz}) is changed only by the replacement of $\cos2\theta$ by
$\cos2\phi$, with no change in $\sin p\theta$.
Now we can ask how the region of repulsion depends on the wedge angle $\beta$.

\begin{figure}
\begin{center}
\includegraphics[scale=1]{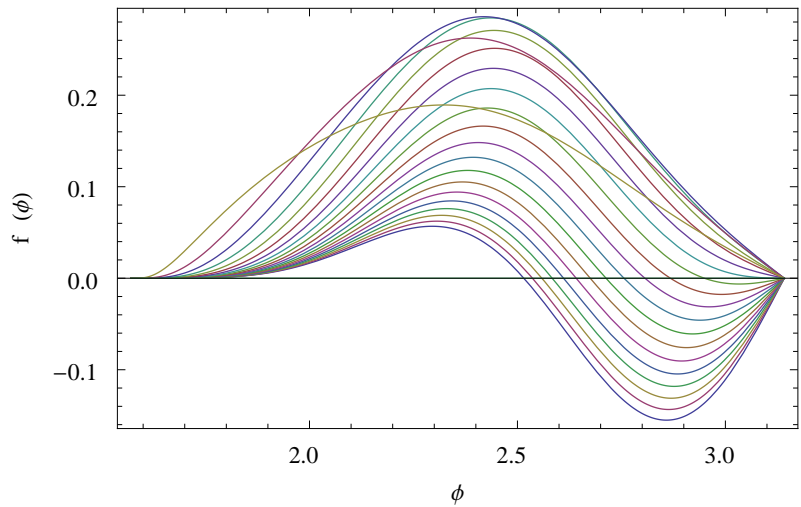}
\caption{\label{fig:cpwedge} The $z$-component of the force on an
completely anisotropic atom moving on a line perpendicular to a
wedge. The different curves are for various values of $\beta$ from
0 to $\pi$ by steps of $\pi/20$, from bottom up.  The last few values
of $\beta$ have a markedly different character from the others.}
\end{center}
\end{figure}

Write for an atom on the line $x=-X$
\be
U^{zz}_{\rm CP}=-\frac{\alpha_{zz}(0)}{8\pi X^4}V(\phi),
\ee
where
\be
\fl V(\phi)=\cos^4\phi\left[\frac{p^4}{\sin^4\frac\pi2\frac{\phi-\beta/2}{\pi
-\beta/2}}-\frac23\frac{p^2(p^2-1)}{\sin^2\frac\pi2\frac{\phi-\beta/2}{\pi
-\beta/2}}+\frac1{45}(p^2-1)(p^2+11)\cos2\phi\right].
\ee
At the point of closest approach,
\be
V(\pi)=\frac1{45}(4p^2-1)(4p^2+11),
\ee
so the potential vanishes at that point only for the half-plane case,
$p=1/2$, as noted above.  The force in the $z$ direction is
\numparts
\bea
F_z=-\frac{\alpha_{zz}}{8\pi}\frac1{X^5}f(\phi),\\
f(\phi)=\cos^2\phi\frac{\partial V(\phi)}{\partial\phi}.
\eea
\endnumparts
\Fref{fig:cpwedge} shows the force as a function of $\phi$ for fixed $X$.
It will be seen that the force has a repulsive region for angles close
enough to the apex of the wedge, provided that the wedge angle is not too
large.  The critical wedge angle is actually
rather large, $\beta_c=1.87795$, or about
108$^\circ$.  For larger angles, the $z$-component of the force exhibits only
attraction.  Of course, the force is zero for $\beta=\pi$ ($p=1$) because then
the geometry is translationally invariant in the $z$ direction.

\section{Repulsion of an atom by a conducting cylinder}
\label{sec:cpcyl}

\begin{figure}
\begin{center}
\includegraphics{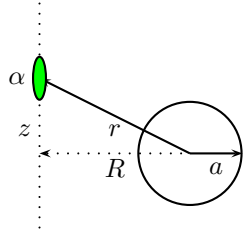}
\caption{\label{fig-cyl-atom} Interaction between an anisotropically 
polarizable
atom and a conducting cylinder of radius $a$.  The force on the atom along 
a line
which does not intersect the cylinder is considered.  If the atom is only
polarizable in that direction, and the line lies sufficiently far from the
cylinder, the force component along the line changes sign near the point of
closest approach.}
\end{center}
\end{figure}

In this section, we consider the interaction of an anisotropic atom
with a perfectly conducting, infinitely long, cylindrical shell.  As above,
we start from (\ref{alphagamma}), and
 assume that the polarizability of the atom has negligible frequency 
dependence
(static approximation), and, in order to maximize the repulsive effect, 
the atom
is only polarizable in the $z$ direction, the direction of the trajectory
(assumed not to intersect the cylinder). Thus, the quantity we need to 
compute for
a conducting cylinder of radius $a$ is given by \cite{bezerra}
\bea
\fl\int_{-\infty}^\infty \frac{\rmd\zeta}{2\pi}\Gamma_{zz}(r,\theta)=
\sum_{m=-\infty}^\infty
\int_0^\infty\frac{\rmd\kappa}{(2\pi)^3}\frac\pi{2a}
\frac1{K_m(\kappa a)K'_m(\kappa a)}
\bigg\{\frac{m^2}{r^2}K_m^2(\kappa r)+\kappa^2K_m^{\prime2}(\kappa r)\nn\\
\quad\mbox{}-\cos2\theta \kappa a[I_m(\kappa a)K_m(\kappa a)]'
\left(-\frac{m^2}{r^2}
K_m^2(\kappa r)+\kappa^2K_m^{\prime 2}(\kappa r)\right)\bigg\}.
\eea
The geometry we are considering is illustrated in \fref{fig-cyl-atom}.
It gives greater insight to give the transverse electric (TE)
and transverse magnetic (TM) contributions to the CP energy:
\numparts
\bea
\fl E^{\rm TE}_{\rm CP}=-\frac{\alpha_{zz}}{4\pi}\sum_{m=-\infty}^\infty
\int_0^\infty d\kappa\,\kappa
\frac{I'_m(\kappa a)}{K'_m(\kappa a)}\left[\frac{\cos^2\theta}{r^2} m^2K_m^2
(\kappa r)
+\kappa^2\sin^2\theta K_m^{\prime 2}(\kappa r)\right],\\
\fl E^{\rm TM}_{\rm CP}=\frac{\alpha_{zz}}{4\pi}\sum_{m=-\infty}^\infty
\int_0^\infty d\kappa\,\kappa
\frac{I_m(\kappa a)}{K_m(\kappa a)}\left[\frac{\sin^2\theta}{r^2} m^2K_m^2
(\kappa r)+\kappa^2\cos^2\theta K_m^{\prime 2}(\kappa r)\right].
\eea
\endnumparts
  The distance of the atom from the center of the
 cylinder is $r=R/\sin\theta$, where $R$ is the distance of
closest approach and $\theta$ is the polar angle, which ranges from
0 when the atom is at infinity to $\pi/2$ when the atom is closest to
the cylinder.

\begin{figure}
\begin{center}
\includegraphics{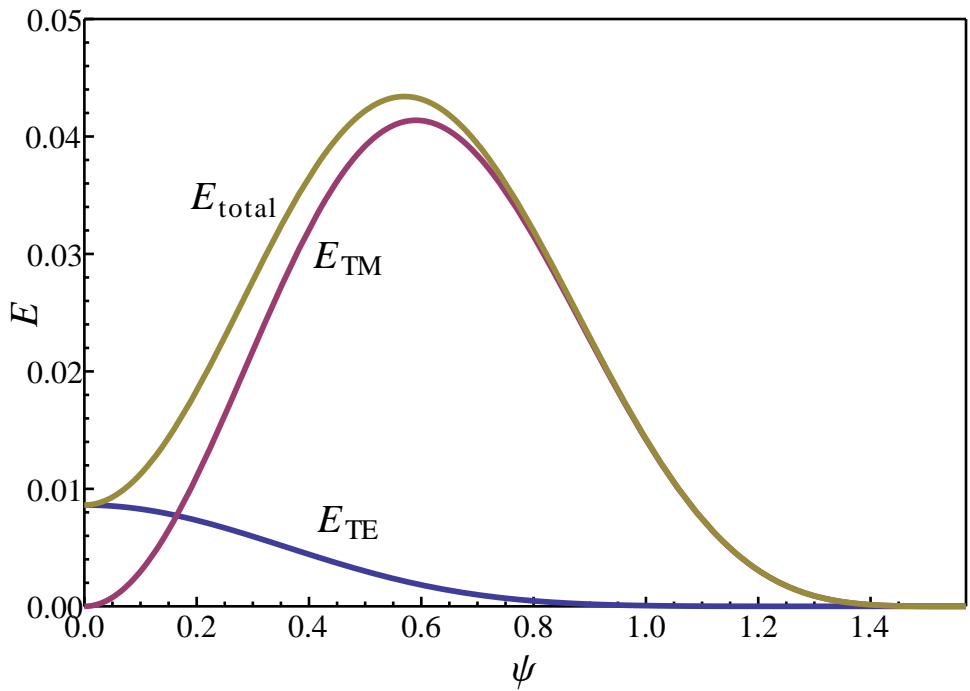}
\caption{\label{fig1} $m=0$ contributions to the Casimir-Polder energy
between an anisotropic atom and a conducting cylinder.  The (generally) lowest 
curve (blue) is the TE contribution, the second (magenta) is the TM 
contribution, and the  top curve (yellow) is the total CP energy.  
In this case, the distance of closest       
approach of the atom is taken to be 10 times the radius of the cylinder. 
The energy $E$ is plotted as a function of $\psi=\pi/2-\theta$.}
\end{center}
\end{figure}

\begin{figure}
\begin{center}
\includegraphics{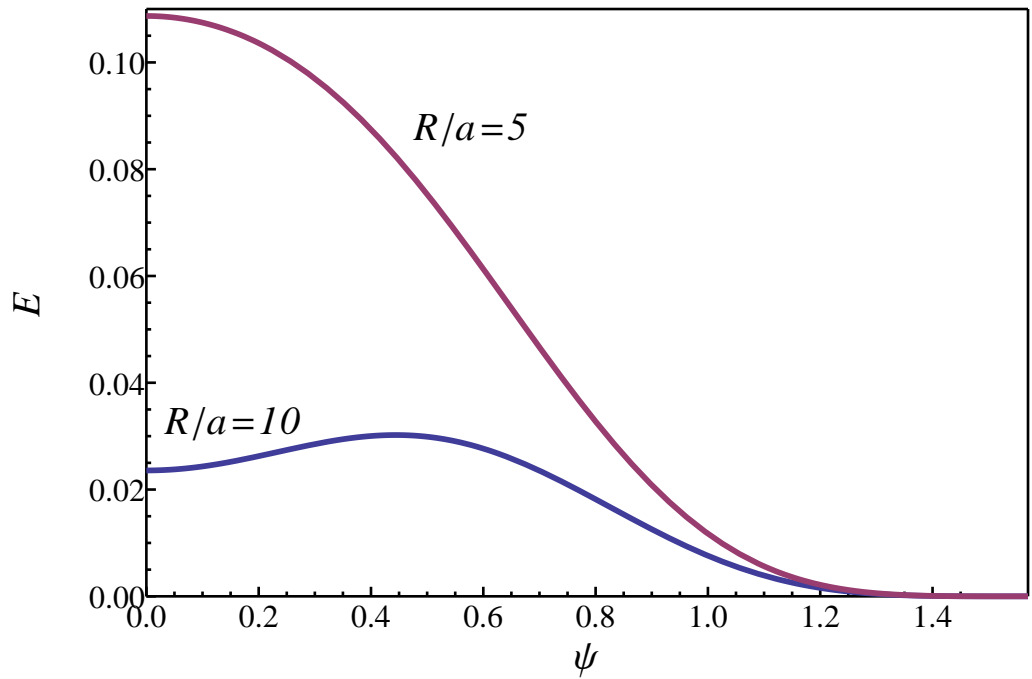}
\caption{\label{figx} The CP energy between an anisotropic atom
and a conducting cylinder.  Plotted is the total CP energy, the upper curve
for the distance of closest approach $R$ being 5 times the cylinder
radius $a$, the lower curve for the distance of closest approach 10 times the
radius.  The curves move up slightly as more $m$ terms are included,
but have completely converged by the time $m=3$ is included.
Repulsion is clearly observed when $R/a=10$, but not for $R/a=5$.}
\end{center}
\end{figure}

At large distances, the CP force is dominated by the $m=0$ term in the
energy sum. \Fref{fig1} shows that for $m=0$ the TM mode dominates
except near the position of closest approach, where only the TE mode
is nonzero.  This indicates that there is a region of repulsion near
$\theta=\pi/2$, since the total energy has a minimum for small
$\psi=\pi/2-\theta$. This effect is partially washed out by including 
higher $m$ modes,
as seen in \fref{figx}, which shows the effect of including the first 5
$m$ values.  But the repulsion goes away if the line of motion passes
too close to the cylinder.  Numerically, we have found that to have
repulsion close to the plane of closest approach requires that
$a/R<0.15$.

What about the analagous calculation of the CP force between a polarizable
atom and a conducting sphere?  Because the latter is symmetric, it is
obvious that there can be no repulsion at large distances, because then
this is the CP interaction between one anisotropic atom and an 
isotropic one.  In fact, numerical calculation reveals that there is
no repulsive regime for even a completely anisotropic atom and a conducting
sphere at any separation distance, as discussed in more detail in  \cite{rep2}.

\section{Conclusions and outlook}
\label{concl}
We have surveyed some of the older and recent work on situations in
which both classical and quantum repulsion between polarizable objects
can occur, with particular stress on the analytic work carried out by
our group.  This is a subject of great current interest, with
many new results emerging, so this brief overview can hardly be definitive.
Especially interesting would be experimental verification
of some of these effects; one might suppose that Rydberg atoms in a high
$l$ state would be sufficiently anisotropic to experience repulsion by a
suitably structured substrate.  However, this now seems unlikely, since
it is very difficult to achieve an anisotropy greater than 1/3.  A Rydberg
atom may be very oblong, but its chief transitions are to more isotropic
states.  This will be discussed in more detail in a forthcoming paper.

\ack
This paper is dedicated to Stuart Dowker, whose contributions to
the subject of quantum vacuum energy are legendary and pervasive.
We thank the US National Science Foundation, the US Department of 
Energy, and the Julian Schwinger Foundation  for 
the support of this work.  We acknowedge the contributions of our colleagues,
S. Fulling, M. Schaden, and K.V. Shajesh, for
their collaborative assistance.


\end{document}